\documentstyle[aps,prb,preprint]{revtex}
\begin{document}
\title{Lattice instabilities of cubic NiTi from first principles} 

\author{Xiangyang Huang, Claudia Bungaro, Vitaliy Godlevsky and Karin M. Rabe}
\address{
Department of Physics and Astronomy, Rutgers University, 
Piscataway, NJ 08854-8019}

\date{\today}
\maketitle
\begin{abstract}
The phonon dispersion relation of NiTi in the simple cubic {\it B}2
structure is computed using first-principles density-functional 
perturbation theory with pseudopotentials and a plane-wave basis
set. Lattice instabilities are observed to occur
across nearly the entire Brillouin zone, excluding three interpenetrating
tubes of stability along the (001) directions and small spheres
of stability centered at R.
The strongest instability is that of the doubly degenerate M$_{5'}$ mode. 
The atomic displacements of one of the eigenvectors of this mode
generate a good approximation to the observed {\it B}19' ground-state structure. 

\end{abstract}


\section{Introduction}
The realization of the technological promise of
active materials in general, and shape memory alloys in particular, 
requires significant progress in the fundamental understanding of
their nature and behavior.
The relative chemical and structural simplicity of the shape
memory alloy NiTi has made it a popular subject of both experimental
and theoretical study\cite{James:2000}.

The monoclinic {\it B}19' low-temperature structure of NiTi is related to
the cubic {\it B}2 high-temperature structure by approximately rigid shifts of alternate
(110) planes along the (1${\bar 1}$0) direction, resulting in a lowering
of the symmetry and a corresponding change in the 
shape of the unit cell\cite{Hehemann:1971}. 
Previous first-principles studies have focused on accurate 
prediction of the ground-state structural parameters by relaxing forces
and stresses for a particular choice of space group, complementing 
the experimental structural determination data\cite{Ye,Sanati}.

The distortion that relates the two structures has the symmetry of a
single M-point normal mode of the {\it B}2 structure.
This suggests a parallel between the martensitic transformation from the 
{\it B}2 to the {\it B}19' structure in NiTi and the paraelectric-ferroelectric 
``soft-mode" transition in perovskite oxides
such as PbTiO$_3$, BaTiO$_3$ and KNbO$_3$\cite{Lines:77}. 
In these perovskites, first-principles calculation of
the eigenfrequencies and eigenvectors of the unstable phonons of the 
cubic high-symmetry prototype structure \cite{Ghosez,Krakauer,Waghmare} 
has shown that an excellent approximation to the ferroelectric ground-state structure is obtained 
by ``freezing in" the single dominant unstable phonon (a polar optic mode 
at the zone center) with an accompanying change in the shape of the unit cell. 
The q-dependence of the instability across the Brillouin zone has been
shown to determine the nature of local structural fluctuations in the
high-temperature cubic paraelectric phase\cite{Krakauer} and, through a first-principles
effective Hamiltonian analysis, the transition temperature and related
quantities\cite{Waghmare}.

In this paper, we take the first step in extending this approach to
shape memory alloys by computing the phonon eigenfrequencies and
eigenvectors of the cubic high-symmetry prototype structure of NiTi
from first principles, as described in Section II.
In Section III, the dominant unstable phonon is identified and the 
structure it generates
is compared with available knowledge of the ground state structure.
We then investigate how these instabilities extend across the 
Brillouin zone. We conclude by speculating on the implications of this
observation on the structure and properties of the high-temperature phase and 
on the martensitic transition.

\section{Computational Method}
First-principles calculations of the structural energetics of {\it B}2 NiTi were
carried out within density-functional theory with a plane-wave pseudopotential
approach. Phonon eigenfrequencies and
eigenvectors throughout the Brillouin zone were obtained using the Green's function
formulation of density-functional perturbation theory (DFPT)\cite{Baroni}.
The calculations were performed with 
the PWSCF and PHONON codes \cite{SISSAcode}, using
the Perdew-Zunger\cite{Perdew:81} 
parametrization of the local-density approximation (LDA).  
Ultrasoft pseudopotentials \cite{Vanderbilt}
for Ti and Ni were generated according to a modified
Rappe-Rabe-Kaxiras-Joannopoulos (RRKJ) scheme \cite{Rappe} 
with three Bessel functions \cite{DalCorso}. 
The electronic wave functions were represented in a
plane-wave basis set with a kinetic energy cutoff of 30 Ry.
The augmentation charges were expanded up to 660 Ry.
The Brillouin zone (BZ) integrations were carried out by the 
Hermite-Gaussian smearing technique \cite{Methfessel} using a 56 k-point mesh 
(corresponding to $12\times12\times12$ regular divisions along 
the $k_x$, $k_y$ and $k_z$ axes) in the $1\over48$ irreducible 
wedge. The value of the smearing parameter was $\sigma$=0.02 Ry. 
These parameters yield  phonon frequencies converged within 
5 cm$^{-1}$. The dynamical matrix was computed on a 
$6\times6\times6$ q-point mesh commensurate with
the k-point mesh. The complete phonon dispersion relation was 
obtained through the computation of real-space interatomic
force constants within the corresponding box \cite{Giannozzi}.

In Table \ref{table1}, we report the equilibrium lattice parameter
and elastic constants of {\it B}2 NiTi obtained
from pseudopotential total energy calculations performed as described above,
and from a previous pseudopotential calculation using a mixed basis set(MB)\cite{Ye}.
For comparison, we also performed full-potential 
linearized-augmented-plane-wave calculations (FLAPW) within both 
the LDA and the generalized-gradient approximation (GGA)\cite{FLAPW}. 
The lattice parameter and bulk moduli were obtained from a Birch function fit \cite{Birch}.
The discrepancy between the PWSCF and FLAPW values for the LDA lattice parameter
is comparable ($<1\%$) to that between the two pseudopotential results.
The FLAPW lattice parameter is slightly larger within GGA, as expected, and the bulk moduli 
show the typical decrease with increasing lattice parameter. 
Although these results are not directly comparable with the properties of the
entropically-stabilized high-temperature {\it B}2 phase, characterized by
large fluctuating atomic displacements, the experimental data included in Table 
\ref{table1} are in general agreement with the first-principles values.
The DFPT calculations reported in the next section were performed at
the PWSCF lattice parameter 5.594 a.u.

\section{Results and Discussion}

The full phonon dispersion of {\it B}2 NiTi consists of six branches: 
three acoustic and three optic. 
Results of our calculations along the 
high-symmetry lines of the simple cubic BZ are shown in
Figure~\ref{phonon}. The imaginary frequencies of the 
unstable modes are represented as negative values. 
The values of the phonon frequencies at the symmetry points are also listed
in Table \ref{table2}.
All of the eigenvectors at the symmetry points $\Gamma$, X, M and R are uniquely
determined by symmetry considerations except for the M$_{5'}$
modes, to be discussed in more detail below.

As expected on the basis of experimental observation of the
low-temperature structure, the {\it B}2 structure 
is unstable at T = 0.
The dominant instability is a doubly-degenerate M$_{5'}$ mode. 
Since there is a second M$_{5'}$ mode (at high frequency), the atomic 
displacements corresponding to the unstable M$_{5'}$ mode are 
not uniquely determined by symmetry, but can be obtained only by
diagonalizing the computed dynamical matrix.
In addition, since the mode is doubly degenerate, there is a two-dimensional
space of phonon eigenvectors that all correspond to
the same frequency.
Two particularly symmetrical eigenvectors in this space are shown in Figure~\ref{eigenvectors}.
In fact, the distorted structure produced by freezing the eigenvector in Figure~\ref{eigenvectors}(a) into the reference cubic structure yields, with an
appropriate choice of overall amplitude, an
excellent approximation to the observed ground-state {\it B}19' structure. 
As can be seen in Table III, the main difference is in the angle $\gamma$,
lowering the symmetry from orthorhombic to monoclinic.
Energetically, this eigenvector is singled out from the others in this space
not at quadratic order, 
but as the result of higher order terms, including strain coupling, not
reported here.
Detailed calculations will be presented in a future publication.

While the evolution of the lattice instability away from the zone center cannot
be directly obtained experimentally, Figure~\ref{phonon} shows that the unstable 
phonon branches actually extend throughout the Brillouin zone.
From the M point, 
the instability extends more than half of the way to $\Gamma$ 
and X. At $\Gamma$, the acoustic branches behave normally at 
very small q.  As q increases along 
$\Gamma$-R and $\Gamma$-M, one and two individual optic modes, 
respectively, disperse strongly downward, mixing with the 
acoustic branches and then becoming unstable. 
As the R point is approached along $\Gamma$-R, 
the unstable longitudinal mode turns around and 
becomes stable once again.
Along $\Gamma$-X the $\Delta_1$ optic mode similarly
disperses downward and mixes with the $\Delta_1$ acoustic branch,
although along this line all phonon modes remain stable.

This behavior can be better visualized in a three-dimensional 
view of the BZ showing the $\omega^2$=0 isosurface 
(Figure~\ref{isosurface}). 
The region of stability is confined to three interpenetrating 
tubes along the cartesian axes, with additional bulges in the 
central region along the $\Gamma$-R directions, and to small 
approximately spherical regions around the R points. 
The unstable modes at M are thus continuously connected to 
the unstable mode along $\Gamma$-R.

Phonon frequencies along symmetry lines in the high-temperature {\it B}2 phase 
have been experimentally measured\cite{Tietz:84} and theoretically analyzed\cite{Zhao:93}.    
As explained in Section II, the properties of this phase are not directly
comparable with calculations for the unstable zero-temperature {\it B}2 structure
(note, however, that the calculated stable acoustic mode frequencies are in general
agreement with their experimental counterparts).
The high-temperature {\it B}2 structure is entropically stabilized and 
should have strong local distortions away from the high-symmetry average positions 
of the atoms. As in the previous work on KNbO$_3$, the nature of these disortions
should be related to the distribution of unstable modes in the BZ, and
could be experimentally characterized through diffuse X-ray scattering or 
with local probes, such as EXAFS.
The soft-mode behavior of NiTi is also less straightforward than that of the
ferroelectric perovskites. In the cubic phase, a soft TA mode at 
$q = {2 \pi \over a_0}({1\over3}{1\over3}0)$
freezes in to generate the intermediate-temperature R phase, and a softening
of the M$_{5'}$ phonon associated with the ground-state B19' phase has not been
directly observed. Theoretical investigation of the phase diagram and 
temperature-dependence of the phonon dispersion requires a first-principles 
effective Hamiltonian analysis, analogous to that for the perovskite oxides,
beyond the scope of the present paper.

\section{Conclusions}

In conclusion, 
we performed {\it ab initio} calculations of the phonon
dispersion of the {\it B}2 high-symmetry reference structure of NiTi. There
are lattice instabilities throughout 
the entire Brillouin zone, with the dominant instability at M. 
A good approximation to the observed ground state structure of NiTi can be 
generated by freezing in a particular choice of eigenvector
from the corresponding two-dimensional space. 

\acknowledgments

We thank R. D. James, K. Bhattacharya and I. I. Naumov for valuable discussions. 
This work was supported by AFOSR/MURI F49620-98-1-0433.
A portion of the calculations were performed on the SGI Origin 2000 at 
ARL MSRC.

%%%%%%%%%%% figure 1 %%%%%%%%%%%
\begin{figure}
  \caption{Phonon dispersion for NiTi in the {\it B}2 structure with 
   a$_0$ = 5.594 a.u. along symmetry lines in the simple cubic BZ. 
   Symmetry labels are assigned according to the conventions of 
   Ref.~\protect\onlinecite{Bassani} with Ni at the origin.
   The imaginary frequencies of the unstable modes are plotted as 
   negative values.}
  \label{phonon}
\end{figure}

%%%%%%%%%%% figure 2 %%%%%%%%%%%
\begin{figure}
  \caption{Relative atomic displacements corresponding to two 
   linearly independent eigenvectors of the unstable $M_{5}^{'}$
   mode at $\vec k = {2 \pi \over a_0}({1\over2}{1\over2}0)$, 
   transforming as 
   (a) $\hat x + \hat y$ and (b) $\hat y$. Here, a portion of 	
   the {\it B}2 structure, with Ni atoms shown by filled circles 
   and Ti atoms shown by open circles, is viewed along the (001)
   direction. All displacements lie in the x-y plane.
}
  \label{eigenvectors}
\end{figure}

%%%%%%%%%%% figure 3 %%%%%%%%%%%
\begin{figure}
  \caption{Zero-frequency isosurfaces of the phonon dispersion 
   relation in the first octant of the Brillouin zone.
   Outside the interpenetrating tubes along the (001) directions 
   and the small spheres centered at R.}
  \label{isosurface}
\end{figure}

%%%%%%%%%%% TABLE I %%%%%%%%%%%
\begin{table}
    \caption{Lattice parameter and elastic properties of NiTi 
             in the {\it B}2 structure.  
             $a_0$ in the parenthesis is GGA result. }
    \begin{tabular}{cccccc}
     &PW&MB(Ref.\ \onlinecite{Ye})&FLAPW\tablenotemark[1]&Exp.\\
    \tableline
  $a_0$(a.u.)&5.594&5.626&5.561(5.670)&5.698(Ref.\ \onlinecite{Goo})\\
  B(Mbar)&1.68&1.56&1.86&1.40(Ref.\ \onlinecite{Mercier})\\
  $c_{11}$(dyn/cm$^2$)&1.80&1.68&-&1.62(Ref.\ \onlinecite{Mercier})\\
  $c_{12}$(dyn/cm$^2$)&1.68&1.44&-&1.29(Ref.\ \onlinecite{Mercier})\\
  $c_{44}$(dyn/cm$^2$)&0.77&0.50&-&0.35(Ref.\ \onlinecite{Mercier})\\
  \tablenotetext[1]{This work. see Ref.\ \onlinecite{FLAPW}}
    \end{tabular}
    \label{table1}
\end{table}
%%%%%%%%%%% END OF TABLE I %%%%

%%%%%%%%%%% TABLE II %%%%%%%%%%%
\begin{table}
    \caption{Computed phonon frequencies of {\it B}2 NiTi with a$_0$ = 5.594 a.u.
             at the symmetry points in the simple cubic BZ.}
    \begin{tabular}{cc|cc}
  Label&Frequency(cm$^{-1}$)&Label&Frequency(cm$^{-1}$)\\
    \tableline
   $\Gamma_{15}$&0    &R$_{15}$&30\\
   $\Gamma_{15}$&200  &R$_{25}$&252\\
   X$_{1'}$&77        &M$_{5'}$&91$i$\\
   X$_{5'}$&121       &M$_{1'}$&91\\  
   X$_{5}$&150        &M$_{4'}$&140\\  
   X$_{1}$&293        &M$_{5'}$&264\\
    \end{tabular}
    \label{table2}
\end{table}
%%%%%%%%%%% END OF TABLE II %%%%

%%%%%%%%%%% TABLE III %%%%%%%%%%%
\begin{table}
    \caption{Structural parameters of {\it B}2 NiTi distorted by freezing in of the
eigenvector in Figure~\ref{eigenvectors},
       {\it a}=2.960 \AA,
       {\it b}=4.186 \AA,
       {\it c}=4.186 \AA,
       $\gamma$=90$^\circ$,
       compared with the experimental monoclinic {\it B}19' structure(Ref.~\protect\onlinecite{KTMO}),
       {\it a}=2.898 \AA,
       {\it b}=4.646 \AA. 
       {\it c}=4.108 \AA,
       $\gamma$=97.78$^\circ$.
While the distorted {\it B}2 structure actually has space group {\it Pmma}, the structure
is presented within $P2_{1}/m$, the space group of the {\it B}19' structure, to facilitate comparison.     }
    \begin{tabular}{lcccc}
    Structure&Wyckoff position&x&y&z\\
    \tableline
    distorted {\it B}2&Ti(2e)&0.5&0.222&0.25\\
       &Ni(2e)&0&0.678&0.25\\
    {\it B}19'&Ti(2e)&0.4176&0.2164&0.25\\
       &Ni(2e)&0.0372&0.6752&0.25\\
    \end{tabular}
    \label{table3}
\end{table}
%%%%%%%%%%% END OF TABLE III %%%%

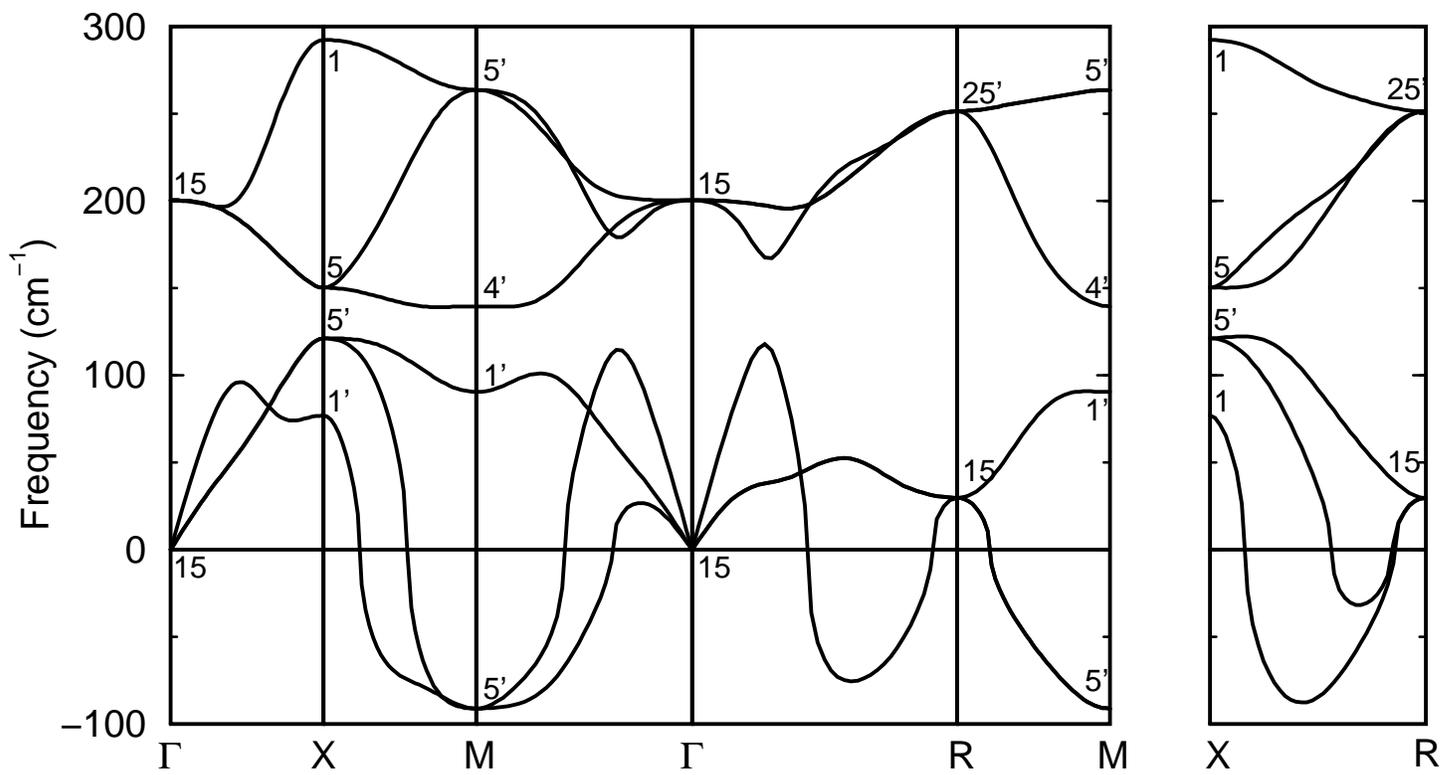
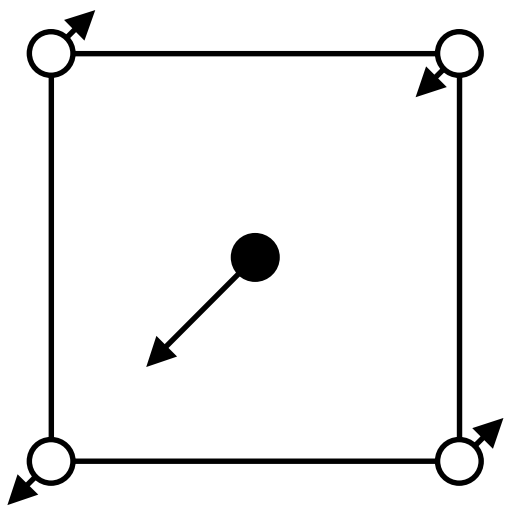
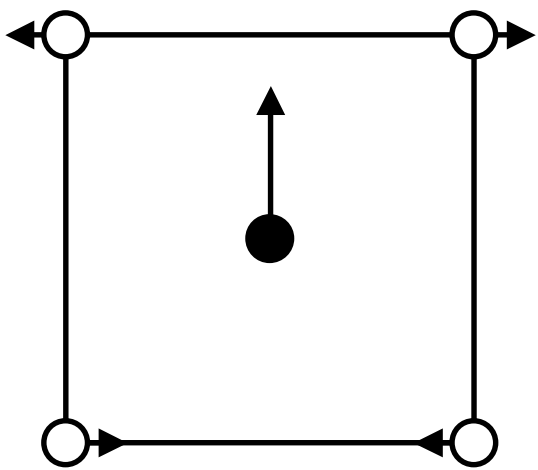
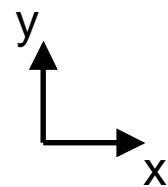
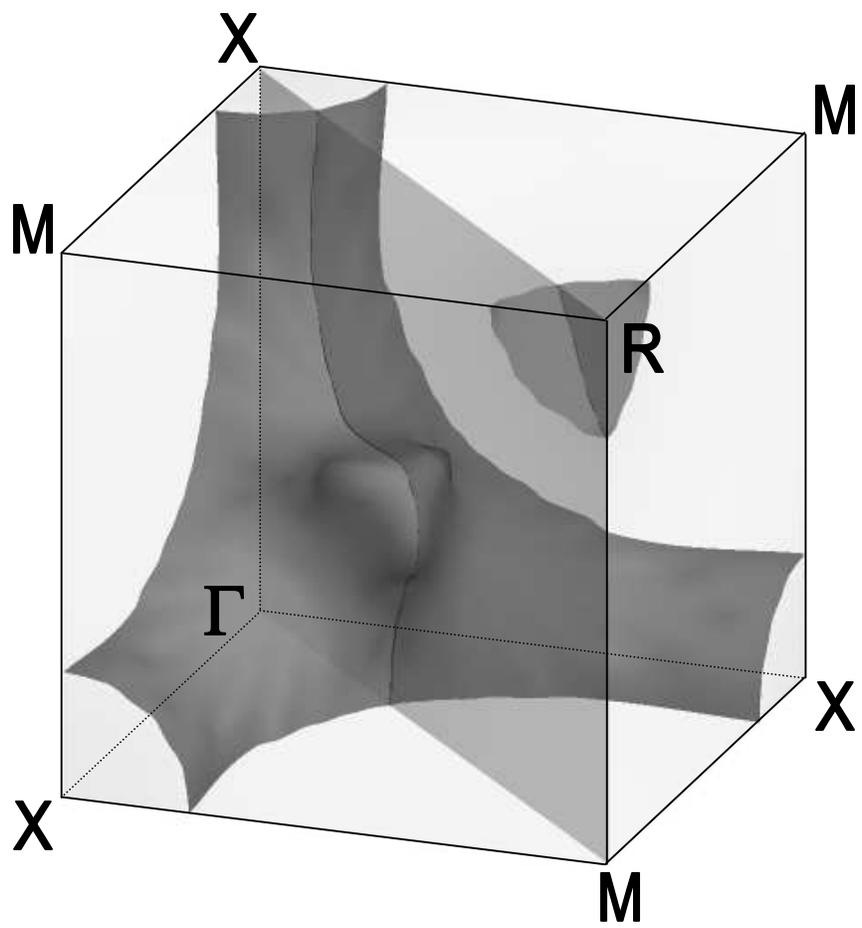
\end{document}